\documentstyle[sprocl]{article}

\input{psfig}
\begin{document}
\vspace*{-7em}
\hfill ADP-01-27/T461 
\vspace{4em}
\title{Distortions in the negative energy Dirac sea:\\ 
violation of the Gottfried sum rule and $\Delta\bar{u}$ in proton}
\author{K. Tsushima$^1$\footnote{Presented at the Joint Workshop of the
Special Research Center for the Subatomic Structure of Matter (CSSM) and
the National Institute for Theoretical Physics on {\it Lepton Scattering,
Hadrons and QCD}, March 26 - April 6, 2001, Adelaide, Australia.}, 
A.W. Thomas$^2$}

\address{Special Research Centre for the Subatomic Structure of Matter\\
and Department of Physics and Mathematical Physics, \\
Adelaide University, Adelaide SA 5005, Australia\\E-mail: 
$^1$ktsushim@physics.adelaide.edu.au, $^2$athomas@physics.adelaide.edu.au} 

\author{G.V. Dunne}

\address{Department of Physics, University of Connecticut,\\
Storrs, CT 06269-3046, USA\\E-mail: dunne@phys.uconn.edu}

\maketitle\abstracts{ 
We report on recent work concerning the effect which the 
change in vacuum structure (negative energy Dirac sea), in the presence
of a confining scalar field, has on the nucleon
structure functions and parton distributions.  
Using the Dirac equation in 1+1 dimensions, we show that 
distortions in the Dirac sea are responsible
for part of the violation of the Gottfried sum rule -- i.e., part of the 
flavor asymmetry in the proton sea. 
Our basic argument is that, even if isospin is an exact symmetry,
the presence of a confining
potential changes the vacuum structure, and inevitably leads to a violation 
of SU(2) flavour symmetry in a hadron with a different number of valence
$u$ and $d$ quarks. The same mechanism also leads to a prediction for 
$\Delta\bar{u}$ and $\Delta\bar{d}$.
}

\section{Introduction}

The study of the violation of the Gottfried sum rule~\cite{Gottfried} 
has attracted considerable interest since its experimental confirmation
by the New Muon Collaboration (NMC)~\cite{NMC}.
At present, there are several possible explanations for it,  
including mesonic (mainly pionic \cite{Thomas:1983fh}) contributions
and the Pauli exclusion principle at the 
quark level~\cite{Signal} -- for recent
reviews see Refs.~\cite{Kumano,Speth,Londergan,Vogt}.

Here we report our recent work~\cite{GTT}, 
on this problem, aimed at putting the phenomenon into a more general
context. Our approach is based on a totally nonperturbative effect,
namely the change in the vacuum structure (including distortion of 
the negative energy Dirac sea) in the presence of a confining potential.
Such changes in vacuum structure have been studied in great detail in
other contexts (including the phenomenon of fractional 
charge~\cite{Jackiw,Vreview,MacKenzie}), since the initial 
work of Jackiw and Rebbi~\cite{Jackiw}.

In the case of the nucleon structure functions, 
self-consistent studies of the vacuum structure 
have been made within a chiral
quark soliton model~\cite{Wakamatsu,Diakonov,Goeke,Weigel}. 
However, it is not easy to see how the change in vacuum structure
is directly reflected in the calculated results.

In this report we will explicitly show 
that the existence of a (quark confining)
potential distorts the negative energy
Dirac sea and inevitably leads to the violation of
the Gottfried sum rule. 
We emphasize that our argument holds
even if charge symmetry and flavor SU(2) symmetry are assumed to be valid.
We also predict $\Delta\bar{u}$ in proton based on the same mechanism.

\section{The Dirac equation in 1+1 dimensions}
Consider massless ($m=0$) fermions in $1+1$ dimensions, in the presence of a
static scalar potential $V(z)$. The fermion spectrum is determined from the
Dirac Hamiltonian, 
\begin{eqnarray}
H\, \varphi(z)&=&\left[-i\alpha \frac{d}{dz}+\beta V(z)\right] \varphi(z),\nonumber\\
&=&\left(\matrix{0 & -\frac{d}{dz}+V(z)\cr \frac{d}{dz}+V(z)&0}\right) \varphi(z)
 = E\varphi(z), 
\label{ham}
\end{eqnarray}
where we have chosen the Dirac matrix representation with $\alpha=\sigma_2$
and $\beta=\sigma_1$. We choose $V(z)$ to be an odd function, with a kink-like
profile such that $V(\infty)=V_0=-V(-\infty)$. Then the spectrum 
is necessarily symmetric about $E=0$, and 
there are continuum thresholds at $E=\pm V_0$.
There is a bound state at $E=0$, but it is irrelevant for our discussion.
Instead, we consider bound states in the gap $|E| (\ne 0) < |V_0|$, 
which occur in pairs symmetrically placed about $E=0$.

We choose the profile for
the scalar potential: $V(z)=V_0\, \tanh z$, where $V_0 > 0$.
Then the $E \ne 0$ bound state wave functions   
with energy $E_n^{(\pm)}=\pm\sqrt{n(2V_0-n)}$, are given by
\begin{eqnarray}
\chi_{b,n}^{(\pm)}(z) = 
N_n\left(\matrix{\sqrt{\frac{n}{2V_0-n}}\,
P_{V_0}^{V_0-n}(\tanh z)\cr 
\pm P_{V_0-1}^{V_0-n}(\tanh z)}\right), 
\label{bound}
\end{eqnarray}
for $n=1,2, \dots [V_0]$, where $[V_0]$ means the greatest integer less than
$V_0$. In Eq.~(\ref{bound}), $P_\nu^\mu(z)$ is the associated Legendre function. 
If $V_0$ is an integer then there are
real (but non-normalizable) threshold states obtained by setting $n=V_0$ in
Eq.~(\ref{bound}). The continuum state wave functions 
are given by
\begin{eqnarray}
\varphi_p^{(\pm)}(z)=N_p \left(\matrix{\sqrt{\frac{V_0-ip}{V_0+ip}}\,
P_{V_0}^{ip}(\tanh z) \cr \pm P_{V_0-1}^{ip}(\tanh z)}\right), 
\label{continuum}
\end{eqnarray}
with energy $E_p^{(\pm)}=\pm\sqrt{V_0^2+p^2}$. 
For simplicity, we choose $V_0=\frac{3}{2}$, where there
is a zero energy state, and two
other bound states with energies $E_1^{(\pm)}=\pm \sqrt{2}$:
\begin{eqnarray}
\chi_{b,1}^{(\pm)}(z)  
= \left(\matrix{\sqrt{2}\tanh z\,{\rm sech}^{1/2} z\cr \pm {\rm
sech}^{1/2} z }\right). 
\label{bound1}
\end{eqnarray} 

These solutions are to be compared to the free Dirac equation  
solutions of positive and negative energies ($E_p = \pm |p|$):
\begin{equation}
u_p(z) = \frac{1}{\sqrt{2}}
\left(\begin{array}{c} -ip/|p| \\ 1 \end{array}\right)\:e^{ipz}, 
\hspace{2em}
v_p(z) = \frac{1}{\sqrt{2}}
\left(\begin{array}{c} ip/|p| \\ - 1 \end{array}\right)\:e^{-ipz}. 
\label{freewf}
\end{equation}

Then, the field operator can be expanded in terms of either 
$\{u_p, v_p, -\infty<p<\infty\}$ or 
$\{\chi_{b,n}, \varphi_p^{(+)}, \varphi_p^{(-)}, -\infty<p<\infty\}$, 
each of which is a complete, orthonormal set~\cite{MacKenzie}:
\begin{eqnarray}
\psi(z) &=& \int_{-\infty}^{\infty} \frac{dp}{2\pi}
\left[ b_pu_p(z) + d^\dagger_pv_p(z) \right]\\
&=& e_n\chi_{b,n} + \int_{-\infty}^{\infty} \frac{dp}{2\pi}
\left[ a_p\varphi_p^{(+)}(z) + c_p^\dagger\varphi_p^{(-)}(z) \right].
\label{field}
\end{eqnarray}
The creation and annihilation operators for the set with $V_0 \ne 0$ satisfy,
\begin{equation}
\{e_n^\dagger,e_m\} = \delta_{nm},\: 
\{a_p^\dagger,a_q\} = 2\pi\delta(p-q),\: 
\{c_p^\dagger,c_q\} = 2\pi\delta(p-q), 
\end{equation}
and all other anticomutators vanish.
The operator sets, e.g., $\{e_n,a_p,c_p^\dagger\}$ and $\{b_p,d_p^\dagger\}$  
can be related using the orthogonality (Bogoliubov transformation):
\begin{eqnarray}
e_n &=& \langle \chi_{b,n}|u_p \rangle b_p +  \langle \chi_{b,n}|v_p
\rangle d_p^\dagger, 
\label{ek}\\
a_k &=& \langle \varphi_k^{(+)}|u_p \rangle b_p + \langle 
\varphi_k^{(+)}|v_p \rangle d_p^\dagger, 
\label{ak}\\
c_k^\dagger &=& \langle \varphi_k^{(-)}|u_p \rangle b_p + \langle 
\varphi_k^{(-)}|v_p \rangle d_p^\dagger, 
\label{ck}
\end{eqnarray}
where the bra-ket notation implies, $\int_{-\infty}^{\infty} \frac{dp}{2\pi}$.

Let us focus on the bound state,  
$|\chi_{b,1} \rangle $, where the wave function is normalized to unity and   
its expression is given by Eq.~(\ref{bound1}).
Using the relation of Eq.~(\ref{ek}) and the property of orthonormality, 
the normalization of the bound state can be written by:
\begin{equation}
1 = \langle \chi_{b,1}|\chi_{b,1} \rangle  =
\int_{-\infty}^{\infty} \frac{dp}{2\pi} 
\left( |\langle \chi_{b,1}|u_p \rangle |^2 + |\langle \chi_{b,1}|v_p
\rangle |^2 \right).
\label{normalization}
\end{equation}

In Eq.~(\ref{normalization}), the term $|\langle \chi_{b,1}|v_p \rangle |^2$ 
($|\langle \chi_{b,1}|u_p \rangle |^2$), 
can be interpreted as the probability density distributions with
which the bound state $|\chi_{b,1} \rangle $ would be filled by the 
negative (positive) energy free states, if the potential were turned 
off suddenly (but not adiabatically).

\begin{figure}[hbt]
\vspace{-0.3cm}
\begin{center}
\hspace{0.1cm}
\psfig{figure=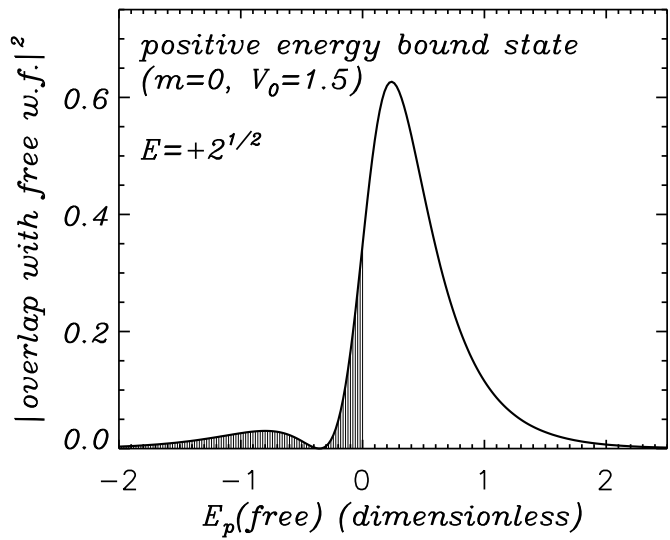,height=8cm}
\vspace{-0.3cm}
\caption{Bound state number density distributions 
as a function of free plane wave energy, $E_p(free)$, 
when projected onto free plane wave solutions.
\label{figboundstate}}
\end{center}
\end{figure}

We show in Fig.~\ref{figboundstate} the quantity 
$|\langle \chi_{b,1}|u_p \rangle |^2 + |\langle \chi_{b,1}|v_p \rangle |^2$,  
(square of the wave function overlaps) as a function of energy for the free plane  
wave solutions, $E_p(free)$.
Surprisingly, one can immediately notice that the positive energy bound  
state wave function, $\chi_{b,1}$, has non-zero 
overlaps with the negative energy free plane wave solutions 
(the shaded area in Fig.~\ref{figboundstate}).
If we integrate over $E_p(free)$, it gives unity by construction, 
or probability conservation, as it should be.
We should also mention that the negative (positive) energy continuum 
state wave functions 
of Eq.~(\ref{continuum}) have also non-zero overlaps with the positive (negative) 
energy free plane wave solutions.
 
The conclusion of this section is that the positive energy bound state,
appearing in the Fock space built on the presence of the potential,
has non-zero overlaps with the negative energy free plane wave solutions.              
It is this point that we want to make a connection with 
the deep-inelastic scattering (DIS)
and the parton model.

\section{Connection with the Parton Model}

In this section we consider the interpretation of the results
obtained in the previous section 
in connection with the parton model.
It may be useful to recall the assumptions which leads to the standard 
parton model according to Ref.~\cite{Jaffe}: 
{\it (1) The current couples to quarks. 
Then the contributions to the forward Compton amplitude can be classified 
by the flow and interactions of quark lines.
(2) At large value of $Q^2$ the currents, but not the states, may be 
treated as in free field theory. Thus, final state interactions and vertex 
corrections are ignored.}

Let us now identify the positive energy bound state wave function of 
Eq.~(\ref{bound1}) as that of the valence quark in a simple, model proton.
We recall that in Fig.~\ref{figboundstate} the quark number density 
distribution is projected onto free plane wave solutions of energy
$E_p(free)$.
As mentioned above, these free plane wave solutions are now to be 
identified with the partons, which appear in free field theory.
As expected, the valence quark wave functions
in the proton have non-zero overlap with the negative energy free plane
waves, and this means that the distributions are non-zero in the
region where Bjorken $x$ is negative. As a result the integral of the
valence distribution over the physical region $x > 0$ will not be one.
\begin{figure}[hbt]
\vspace{-0.3cm}
\begin{center}
\hspace{0.1cm}
\psfig{figure=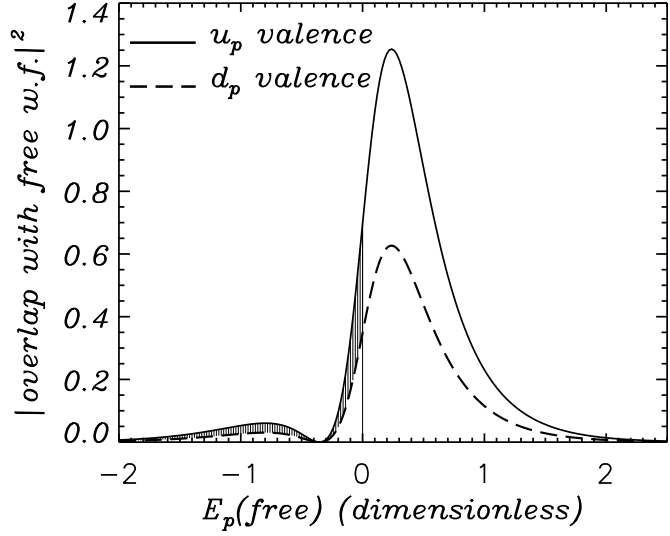,height=8cm}
\vspace{-0.3cm}
\caption{$u$ and $d$ valence quark number density distributions
in the ``proton'' as a function of free energy, $E_p(free)$, 
projected onto free plane waves.
\label{figGottfried}}
\end{center}
\end{figure}
The resolution of this problem lies in the fact that the Dirac sea in
the presence of a scalar potential is also different from that of free
space. In particular, the Dirac sea in the presence of the soliton in
our 1D example (or the confining potential for the proton) consists of
having all negative energy states occupied. If re-expressed in terms of
free plane waves, however, one finds an equal number of positive energy
free states occupied and negative energy states (holes) empty. That is,
when the occupied negative energy states are expressed in terms of free
states one finds an intrinsic sea of $q\bar{q}$ pairs. In the absence of
valence quarks this sea carries no net flavour (or in 3D, spin),
relative to the free space vacuum.

However, once we add a single valence quark (say a $u$ quark), the
negative energy tail will fill some of the holes in the (free space) 
Dirac sea. An equal number of occupied positive energy states from the
$u$ quark Dirac sea will then be interpreted as the missing valence
normalization (because $u_v$ is {\bf defined} to be $u - \bar{u}$).
The net result is that we have one valence $u$ quark, together with a
reduction in the number of $u\bar{u}$ pairs in the sea of the proton.
That is, one will have $\bar{d} > \bar{u}$ by an amount equal to the
area under the valence curve to the left of $E_p = 0$ in Fig. 1. 
If instead we had one valence $u$ and one valence $d$ the intrinsic sea
would be flavour symmetric again, while for two valence $u$'s and one
valence $d$, as in the proton, we have $\bar{d} > \bar{u}$ by the same
amount as for the single valence $u$ (the area under the valence curve
to the left of $E_p = 0$ in Fig. 1).

The situation with respect to the Gottfried sum rule is therefore
illustrated in Fig. \ref{figGottfried}. Because there are more valence
$u$'s, the $u$ sea is depleted more than the sea of $d$'s.
Mathematically, the expression for the Gottfried sum rule is then:
\begin{eqnarray}
S_G &=& \int_0^1 \frac{dx}{x}
\left[ F_2^{\mu p}(x) - F_2^{\mu n}(x) \right], \nonumber\\
&=& \frac{1}{3} \int_0^1 dx
[ u_v(x) - d_v(x)] - \frac{2}{3} \int_0^1 dx [\bar{d}(x) - \bar{u}(x)],
\nonumber\\
&=& \frac{1}{3} - \frac{2}{3} \int_{-\infty}^{\infty} \frac{dp}{2\pi}
|\langle v_p| \chi_{b,1} \rangle |^2.
\label{SG}
\end{eqnarray}
(Note that we have assumed charge symmetry to set $d^n = u^p \equiv u$,
etc.)

\section{Conclusion}
By considering the simple 1D problem outlined here, we aimed to make the
physics origins of one source of flavour asymmetry in the proton
explicit and easy to understand. While one cannot access spin in the 1D
case, the simplicity of the argument for flavour makes the corresponding
argument for $\Delta \bar{u}$ (and $\Delta \bar{d}$) clear. Since in a
spin up proton described by an SU(6) wave function the valence $u$ quark
is predominantly spin up, one will be missing spin up $u$ quarks in the
sea of the proton. Thus $\Delta \bar{u}$ will be positive and
of a similar magnitude to that of $\bar{d} - \bar{u}$ from the same
source \cite{Signal,Cao:2001nu}. Since the $d$ quarks tend to have spin
down in the proton, one also expects $\Delta \bar{d}$ to be negative and
considerably smaller in magnitude than $\Delta \bar{u}$. Since any
contribution to the observed flavour asymmetry from the nucleon's pion
cloud is unpolarized \cite{Thomas:1983fh,Thomas:2000ny}, 
the experimental determination of 
$\Delta \bar{u}$ and $\Delta \bar{d}$ would be extremely valuable in
reaching a quantitative understanding of the roles of these two very
different physical mechanisms in generating the non-perturbative
structure of the proton sea.

\section*{Acknowledgments}
We would like to thank W. Melnitchouk and 
A.W. Schreiber for many helpful discussions.
K.T. would like to acknowledge the hospitality at 
KFA J\"{u}lich, where this report was completed.
This work was supported by the Australian Research Council and Adelaide
University. GD is supported by the US DOE, and thanks the CSSM at Adelaide for
hospitality during a visit in which this work was begun.

\newpage

\end{document}